\documentclass[prl,twoside,showpacs,twocolumn,floatfix]{revtex4}

\usepackage{amssymb}
\usepackage{amsfonts}
\usepackage{graphicx}
\begin{document}

\title{Work fluctuation theorems for harmonic oscillators}
\author{F. Douarche, S. Joubaud, N. B. Garnier, A. Petrosyan, S. Ciliberto}
\affiliation{Laboratoire de Physique de l'ENS Lyon, CNRS UMR 5672, 
        46, All\'ee d'Italie, 69364 Lyon CEDEX 07, France}
\date{\today}

\begin{abstract}

The work fluctuations of an oscillator in contact with a thermostat
and driven out of equilibrium by an external force are
studied experimentally and theoretically within the context of
Fluctuation Theorems (FTs). The oscillator dynamics is modeled by
a second order Langevin equation. Both the transient and
stationary state fluctuation theorems hold 
and the finite time corrections are very
different from those of a first order Langevin equation. The
periodic forcing of the oscillator is also studied; it presents 
new and unexpected short time convergences. 
Analytical expressions are given in all cases.

\end{abstract}


\maketitle
In this letter, we investigate, within the context of the 
\emph{Fluctuation Theorems} (FTs), the work fluctuations of  a
harmonic oscillator in contact with a thermostat and driven out of
equilibrium by an external force.  First found in dynamical
systems~\cite{Evansetal93,GallavottiCohen95a} and later extended to
stochastic systems~\cite{Kurchan98,Farago,Cohen,Cohen1}, these
conventional FTs give a relation between the probabilities to
observe a positive value of the (time averaged) ``entropy
production rate'' and a negative one. This relation is of the form
$P(\sigma)/P(-\sigma) = \exp[\sigma\tau]$, where $\sigma$ and
$-\sigma$ are equal but opposite values for the entropy production
rate, $P(\sigma)$ and $P(-\sigma)$ give their probabilities and
$\tau$ is the length of the interval over which $\sigma$ is
measured. In these systems, the above mentioned FT is derived for
a mathematical quantity $\sigma$, which has a form similar to that
of the entropy production rate in Irreversible Thermodynamics~\cite{Evans:Searles}. 

The proof of FTs is based on a certain number of hypothesis;
experimenting on a real device is useful not only to check those hypothesis, 
but also to observe whether the predicted
effects are observable or remain only a theoretical tool.
There are not many experimental tests of FTs. Some of them are
performed in dynamical systems \cite{otherexperiment} in which the
interpretation of the results is very difficult. Other
experiments are performed on stochastic systems, one on a Brownian
particle in a moving optical trap~\cite{Wangetal:02:05} and another
on electrical circuits driven out of equilibrium by injecting in
it a small current~\cite{Garnier}. The last two systems are
described by first order Langevin equations and the results agree
with the predictions of ref.\cite{Cohen,Cohen1}. As far as we know
no theoretical predictions are available for systems
described by a second  order Langevin equation. The test using
an harmonic oscillator is particularly  important because the
harmonic oscillator is the basis of many physical processes.
Indeed the general predictions of FTs are valid only for $\tau
\rightarrow \infty$ and the corrections for finite $\tau$ have
been computed only for a first order Langevin dynamics.

In the present letter, we address several important questions.
We investigate first the {\em Transient Fluctuation Theorem} (TFT)
of the total external work done on the system in the transient
state, i.e., considering a time interval of duration $\tau$ which
starts immediately after the external force has been applied to
the oscillator. We then analyze the {\em Stationary State
Fluctuation Theorem} (SSFT) which concerns fluctuations in the
stationary state, {\em i.e.}, in intervals of duration $\tau$ starting
at a time long after the external force has been applied. We
also study a new case of stationary behavior obtained when the
system is driven periodically in time~\cite{Zamponi:05}. In this case, which is
actually a very important one, no theoretical prediction is
available. We show that the finite time corrections for SSFT are
already very complex in both these rather simple situations.

To test the FT we measure the
out-of-equilibrium fluctuations of a harmonic oscillator  whose
damping is mainly produced by the viscosity of the surrounding
fluid, which acts as a thermal bath of temperature $T$. We recall here only the main
features of the experimental set-up, more details can be found in
ref.\cite{rsi,DouarcheJSM}. The oscillator is a torsion pendulum
composed by a brass wire and a glass mirror glued in the middle of
this wire. It is enclosed in a cell filled by a water-glycerol
solution at $60\%$ concentration. The motion of this pendulum can
be described by a second order Langevin equation:
\begin{equation}
I_{\mathrm{eff}}\,\frac{{\rm d}^2{\theta}}{{\rm d}t^2} 
+ \nu \,\frac{{\rm d}{\theta}}{{\rm d}t}  + C\,\theta =
M + \sqrt{2k_BT} \eta, \label{eqoscillator}
\end{equation}
where $\theta$ is the angular displacement of the pendulum,
$I_{\mathrm{eff}}$ is the total moment of inertia of the displaced
masses, $\nu$ is the oscillator viscous damping, $C$ is the elastic
torsional stiffness of the wire, $M$ is the external torque,
$k_B$ the Boltzmann constant and
$\eta$ the noise, delta-correlated in time. In our system the measured parameters are
the stiffness $C = 4.5 \times 10^{-4} \textrm{N\,m\,rad}^{-1}$, 
the resonant frequency $f_o=\sqrt{C/I_{\rm eff}}/(2\pi)=217$Hz and 
the relaxation time $\tau_\alpha^{-1}=2I_{\rm eff}/\nu= 9.5$ms. 
The external torque $M$ is applied by means of a tiny electric current $J$ flowing in a
coil glued behind the mirror. The coil is inside a static magnetic
field, hence $M\propto J$. The measurement of $\theta$ is
performed by a differential interferometer, which uses two  laser beams
impinging on the pendulum mirror~\cite{rsi,DouarcheJSM}. 
The measurement noise is two
orders of magnitude smaller than the thermal fluctuations of the
pendulum. $\theta(t)$ is acquired with a resolution
of 24 bits at a sampling rate of $8192$Hz, which is
about 40 times $f_o$. The calibration accuracy of the apparatus,
tested at $M=0$ using the the Fluctuation Dissipation Theorem, is
better than $3\%$(see~\cite{DouarcheJSM}).

\begin{figure}
\centerline{\includegraphics[width=1.0\linewidth]{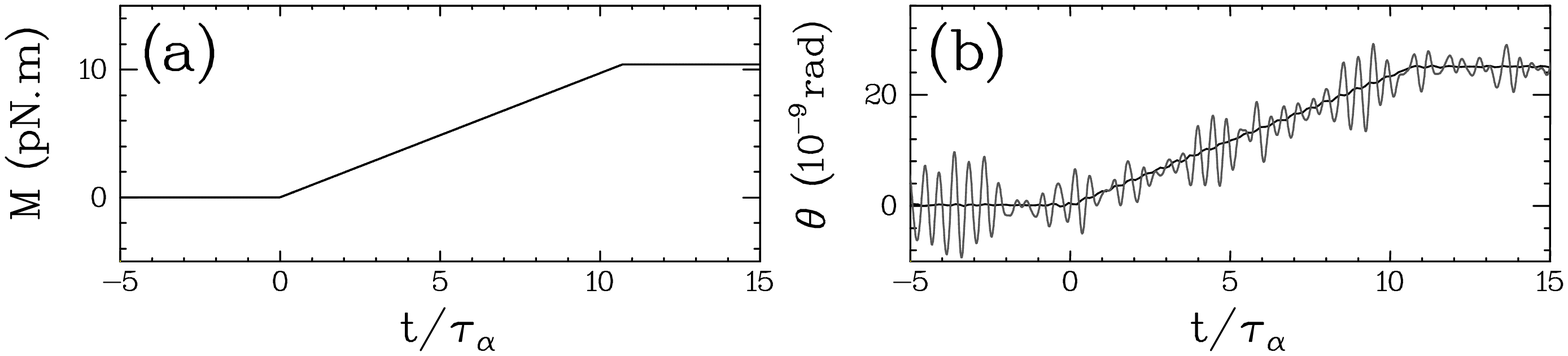}}
\caption{ a) Typical driving torque applied to the oscillator; b)
 Response of the oscillator to the external torque (gray line).
 The dark line represents the mean response $\bar \theta(t)$
 to the applied torque $M(t)$.}
\label{driver_fluct}
\end{figure}

To study SSFT and TFT we apply to the oscillator a time dependent
torque $M(t)$ as depicted in Fig.~\ref{driver_fluct}a, and we 
consider the work $W_\tau$ done by $M(t)$ over a time $\tau$:
\begin{equation}
W_\tau= {1 \over k_B \ T} \ \int_{t_i}^{t_i+\tau} \left[ M(t)-M(t_i) \right]
{d \theta \over dt} dt \,.
\label{work}
\end{equation}
The TFT implies that $t_i=0$ whereas $t_i \ge 3\tau_\alpha$ for SSFT. 
As a second choice for $M(t)$, the linear ramp with a rising time
$\tau_r$ is replaced by a sinusoidal forcing; this
leads to a new form of stationary state which has never been considered in the
context of FT. We examine first the linear forcing $M(t)={M_o t / \tau_r}$ 
(Fig.\,\ref{driver_fluct}a)), with $M_o=10.4$ pN.m and $\tau_r$ = 0.1 s = 10.7 $\tau_\alpha$. The
response of the oscillator to this excitation is comparable to the
thermal noise amplitude, as can be seen in
Fig.\,\ref{driver_fluct}b) where $\theta(t)$ is plotted during the
same time interval of Fig.\,\ref{driver_fluct}a). Because of
thermal noise the power $W_\tau$ injected into the system (eq.\ref{work})
is itself a strongly fluctuating quantity.

\begin{figure}
\centerline{\includegraphics[width=1.0\linewidth]{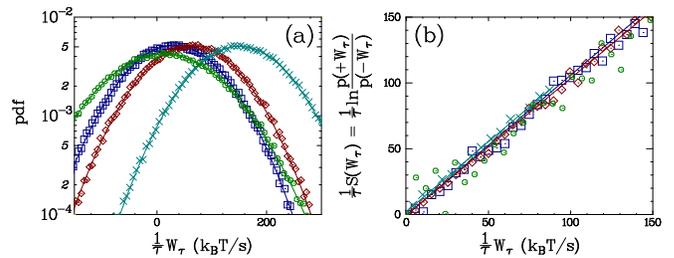}}
\caption{TFT.
   a) $P(W_\tau)$ for TFT for various $\tau/\tau_\alpha$: 0.31 $(\circ)$, 
   1.015 $(\Box)$, 2.09 $(\diamond)$ and 4.97 $(\times)$. Continuous lines
   are Gaussian fits.
   b) TFT; $\S(W_\tau)$ computed with the PDF of a).
   The straight continuous lines are fits with slope 1, {\em i.e.}, $\Sigma(\tau)=1, \ \forall \tau$.
   }
\label{fig:TFT}
\end{figure}

We consider first the TFT. The probability density functions (PDF)
$P(W_\tau)$ of $W_\tau$ are plotted in Fig.\,\ref{fig:TFT}a) for
different values of $\tau$. We see that the PDF are Gaussian for
all $\tau$ and the mean value of $W_\tau$ is a few $k_B T$. We
also notice that the probability of having negative values of $W$ is
rather high for the small $\tau$. The function
\begin{equation}
S(W_\tau) \equiv  \ln\left[{P(W_\tau)\over P(-W_\tau)}\right] \label{eq_FT}
\end{equation}
is plotted in fig.\ref{fig:TFT}b). It is a linear function of $W_\tau$ for any
$\tau$, that is $S(W_\tau)=\Sigma(\tau)~W_\tau$. 
Within experimental error, we measure the slope $\Sigma(\tau)=1$.
%
Thus for our harmonic oscillator the TFT is verified for any time $\tau$. 
This was expected~\cite{Evans:Searles,Cohen1}, and we 
give a derivation of this generic result for a second order 
Langevin dynamics at the end of the letter. 

\begin{figure}
\centerline{\includegraphics[width=1.0\linewidth]{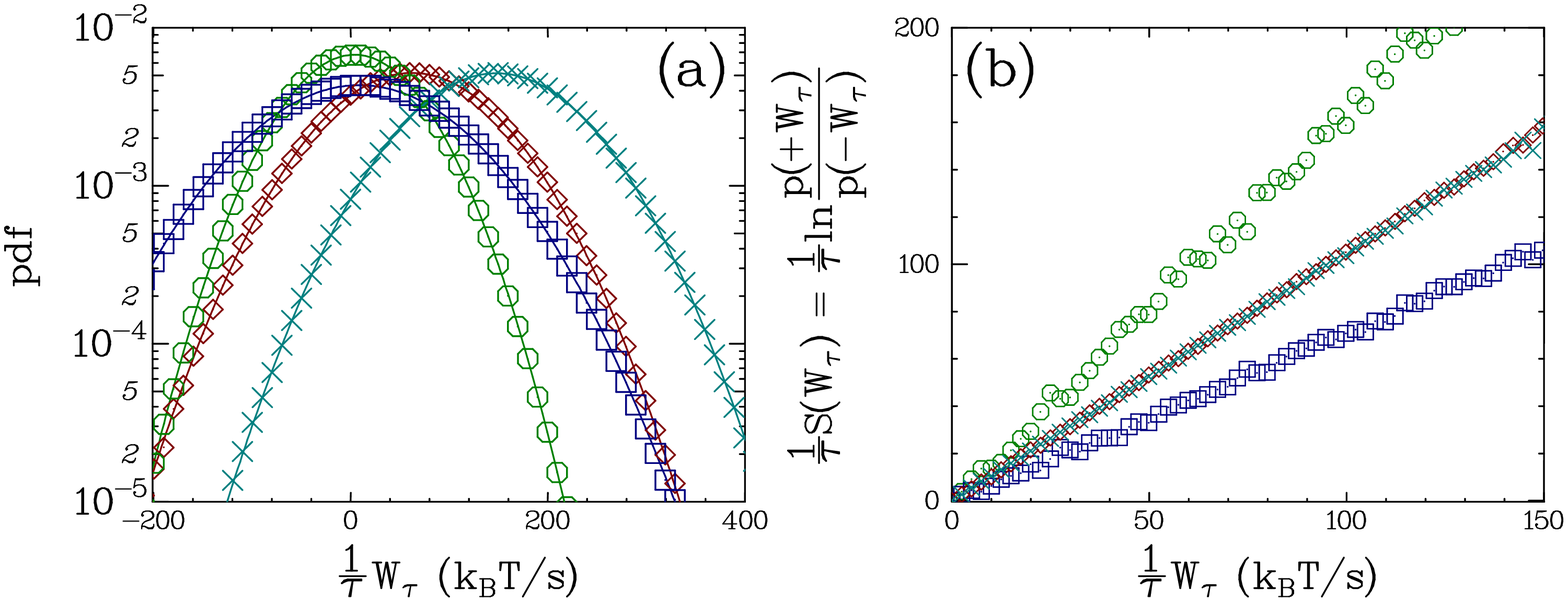}}
\centerline{\includegraphics[width=0.8\linewidth]{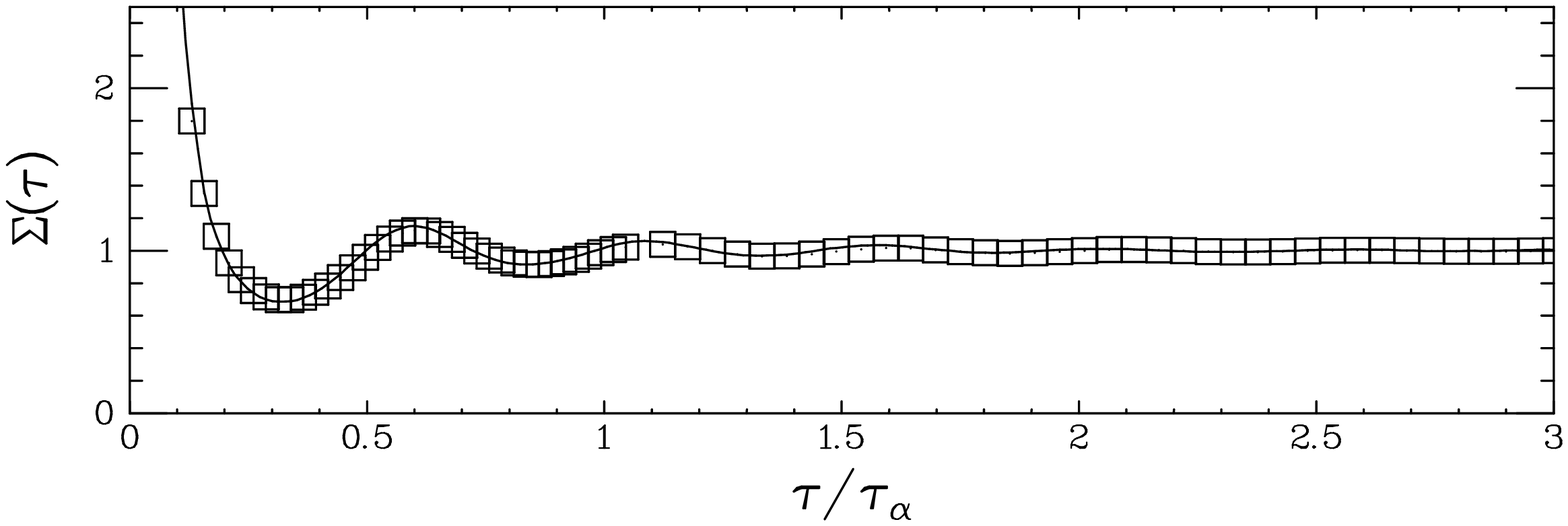}}
\caption{SSFT with a ramp forcing. 
a) PDF of $W_\tau$ for various $\tau/\tau_\alpha$: 0.019 $(\circ)$, 
0.31 $(\Box)$, 2.09 $(\diamond)$ and 4.97 $(\times)$. 
b) Corresponding functions $S(W_\tau)$. 
c) The slope $\Sigma(\tau)$ of $S(W_\tau)$ is plotted
versus $\tau$ ($\Box$: experimental values; continuous line: 
theoretical prediction eq.(\ref{eqSigmaSSFT}) with no adjustable parameter).}
\label{fig:SSFT}
\end{figure}

We now consider the SSFT with $t_i \ge 3\tau_\alpha$ in eq.\ref{work}.
We find that the PDF of $W_\tau$, plotted in Fig.\ref{fig:SSFT}a),
are Gaussian with many negative values of $W_\tau$ for short $\tau$.
The function $S(W_\tau)$, plotted in Fig.\ref{fig:SSFT}b), is
  still a linear function of $W_\tau$, but, in contrast to TFT,  the slope
  $\Sigma(\tau)$ depends  on $\tau$. In Fig.\ref{fig:SSFT}c) the
  measured values of $\Sigma(\tau)$ are plotted as a function of
  $\tau$. The function $\Sigma(\tau)\rightarrow 1$ for
  $\tau \gg \tau_\alpha$. Thus SSFT is verified only for large $\tau$.
The finite time corrections of SSFT, which present oscillations
whose frequency is close to $f_0$, agree quite well with the
  theoretical prediction computed for a second order Langevin
  dynamics that we will discuss at the end of the paper
  (see eq.\ref{eqSigmaSSFT}).
   We stress that the  finite time correction is in
  this case very different from that computed in ref.\cite{Cohen1,Cohen}
  for the first order Langevin equation. 
  
The results of Figs.\ref{fig:TFT},\ref{fig:SSFT} have been checked for
several $M_o/\tau_r$ without noticing any difference. The errors
bars in the figure are within the size of the symbols, and they 
come only from the calibration errors of the harmonic oscillator parameters,
and statistics of realisations (typically $5\times10^5$ cycles have been used).

\begin{figure}
\centerline{\includegraphics[width=1.0\linewidth]{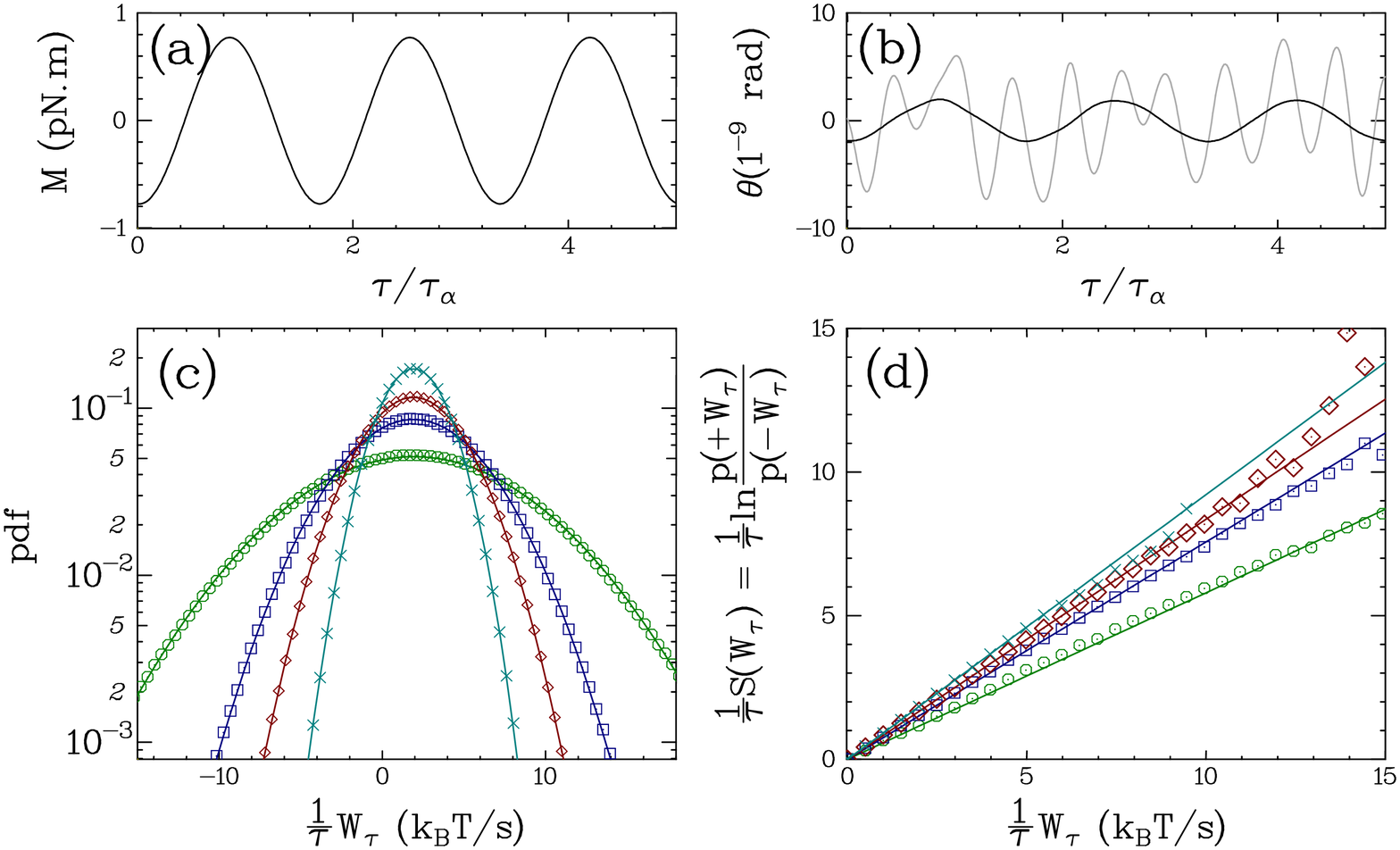}}
\centerline{\includegraphics[width=0.8\linewidth]{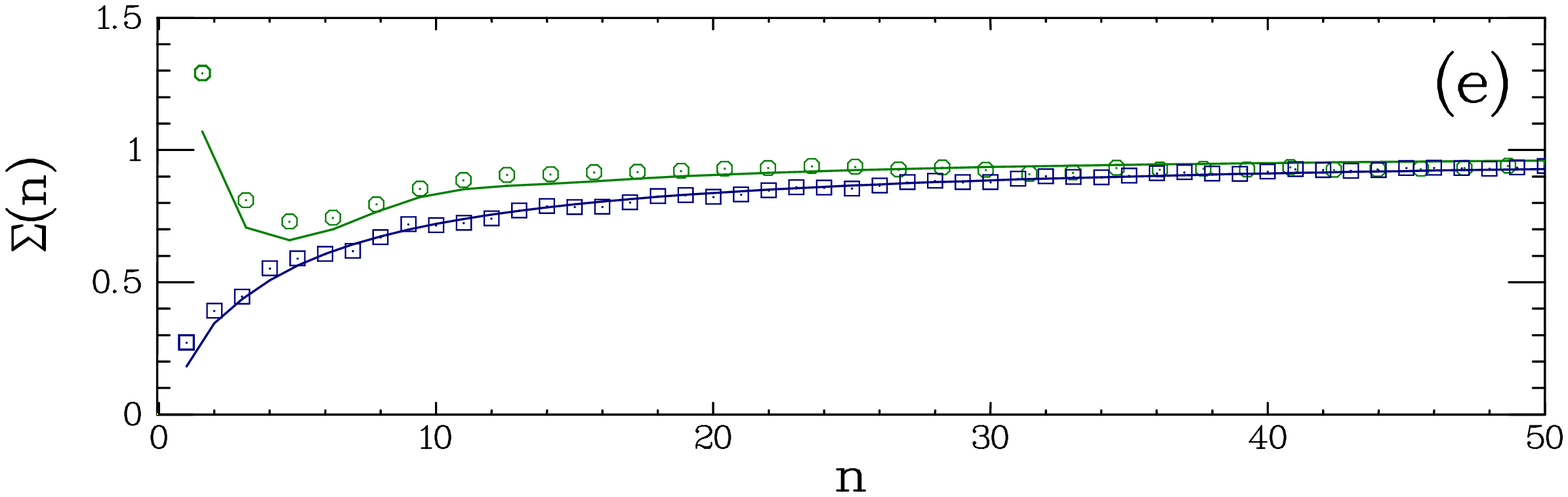}}
\caption{Sinusoidal forcing.
a) Sinusoidal driving torque applied to the oscillator.
b) Response of the oscillator to this periodic forcing (gray line);
 the dark line represents the mean response $\bar{\theta}(t)$.
c) PDF of the work $W_n$ integrated over $n$ periods of forcing, with
$n=7$ ($\circ$), $n=15$ ($\Box$), $n=25$ ($\diamond$) and $n=50$ ($\times$).
d) The function $S(W_n)$ measured at $\omega_d/2\pi=64$Hz is plotted
as a function of $W_n$ for several $n$: $(\circ) n=7$;
$(\Box) n=15$ $(\diamond) n=25$; $(\times) n=50$. Continuous lines are functions $S(W_n)$
computed from Gaussian fits of PDF (in Fig.b).
e) The slopes $\Sigma(n)$, plotted
as a function of $n$ for two different driving frequencies $\omega_d$ = 64 Hz ($\Box$)
and 256 Hz ($\circ$); continuous lines are theoretical predictions from eq.\ref{eqepSINUS}
with no adjustable parameter.} 
\label{fig:sinus}
\end{figure}

Finally we want to briefly describe the results of the periodic
forcing. In this case $M(t)= M_o \sin \omega_d t$ and the work expression (eq.{\ref{work}}) 
is replaced by 
\begin{equation}
W_n = W_{\tau=\tau_n} = {1 \over k_B \ T} \ \int_{t_i}^{t_i+\tau_n} M(t) {d \theta \over dt} dt \,,
\label{eq:work:sinus}
\end{equation}
with $\tau_n=n 2 \pi /\omega_d$ with $n$ integer. 
This is a stationary state that has never been studied before 
in the context of FT. We find that for any driving frequency
$\omega_d$ the PDF of $W_n$ are Gaussian. The function
$S(W_n)$, measured at $\omega_d/2\pi=64$Hz and plotted in
Fig.\ref{fig:sinus}a), is linear in $W_n$ and the corresponding
slope $\Sigma(n)$ is a function of $n$. The measured values
of $\Sigma(n)$ are shown  as function of $n$ in
fig.\ref{fig:sinus}b), where the results obtained at $\omega_d/2\pi=256$Hz
are plotted too. We see that the convergence rate is quite
different in the two cases, which agree with our theoretical
predictions for a second order Langevin equation (see
eq.\ref{eqepSINUS}).
Also in the case of the sinusoidal forcing the agreement between
the computed and measured finite time corrections is very good.
These results prove not only that FTs asymptotically hold for any kind of
forcing, but also that finite time corrections strongly depend on the
specific dynamics. 
In the case of the sinusoidal forcing, the convergence is very slow: in Fig.~\ref{fig:sinus}e), 
we see that it takes several dozens of excitation period (500 ms for $\omega_d/2\pi=64$Hz) to get 
$\Sigma(n)=1$ within one percent. On the contrary for a ramp forcing, this was achieved after 
a few $\tau_\alpha$ (20 ms), see Fig.~\ref{fig:SSFT}c).

\begin{figure}
\centerline{\includegraphics[width=1.0\linewidth]{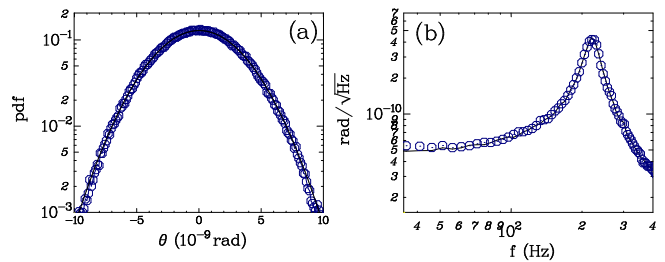}}
\caption{a) PDF of the fluctuations $\delta
\theta= \theta -\bar \theta(t)$ when the torque is applied ($\circ$), 
compared with a Gaussian fit of the PDF at equilibrium (continuous line). 
b) The measured spectrum of $\delta \theta$ ($\circ$) is compared with the prediction
of fluctuation dissipation theorem in equilibrium (continuous line). }
\label{fluctuations}
\end{figure}

Let us now compute the finite time corrections to TFT and SSFT (plotted in
Figs.\ref{fig:SSFT}c,\ref{fig:sinus}e) for the harmonic
oscillator, applying a method very similar to the one
already used in the the context of the Jarzynski equality (see
ref.\cite{DouarcheJSM}). We write $\theta(t)=\bar \theta(t)+
\delta \theta(t)$ where $\bar \theta(t)$ is the mean response of
the system to the external torque and $\delta \theta(t)$ are the
thermal fluctuations. The mean response $\bar \theta (t)$
(dark line in Fig.\,\ref{driver_fluct}b resp. \ref{fig:sinus}b) is computed  by
performing an ensemble average of $\theta(t)$ over $10^3$
responses to the $M(t)$ of Fig.~\ref{driver_fluct}a) resp. Fig.~\ref{fig:sinus}a). It turns
out that the measured $\bar \theta(t)$ is equal to the solution of
eq.\ref{eqoscillator} with $\eta=0$ and with $M$ equal to the
applied time dependent torque. Once the mean behavior is known, it
is useful to compare the statistical properties of $\delta \theta(t)$
measured at $M(t)\ne 0$ with the equilibrium ones. In Fig.\ref{fluctuations}a)
we plot the gaussian fit of the PDF of $\delta \theta(t)$ measured 
at equilibrium $M(t)=0$ (continuous line) and at $M(t)\ne0$
($\circ$). The  two curves are equal
 within experimental errors. Thus we  conclude that the external
 driving does not perturb the equilibrium PDF,
  which is a Gaussian of variance $k_B T/C$.
  In Fig.\ref{fluctuations}b) we plot the power spectra of $\delta
\theta$ in equilibrium (continuous line) and  out  equilibrium
(dotted line). The two spectra are equal  and they coincide with
the theoretical spectrum of an equilibrium second order Langevin
dynamics (eq.\ref{eqoscillator} with $M=0$) computed using the
oscillator parameters. Thus we clearly see that, within
statistical accuracy, the fluctuations of $\delta \theta$ measured
at $M(t)\ne 0$ are those of equilibrium. This important
observation is the key point to estimate the finite time
corrections of FTs.

In order to compute $\Sigma(\tau)$, we first decompose the total
work of the external torque into the sum of a mean part and a 
fluctuating one, {\em i.e.} $W_\tau= \bar{W}_\tau + \delta W_\tau$, 
where
\begin{equation}
\bar{W}_\tau = \frac{1}{k_B T} \int_{t_i}^{t_i+\tau} \left[ M(t)-aM(t_i) \right] 
{d \bar \theta \over dt} dt \,,
\label{barW}
\end{equation}
%
%
\begin{equation}
\delta W_\tau= \frac{1}{k_BT} \int_{t_i}^{t_i+\tau} [M(t)-aM(t_i)] {d
\delta \theta \over dt} dt \,,
\label{eqdeltaW}
\end{equation}
with $a=1$ for the ramp forcing (eq.~\ref{work}) 
or $a=0$ for sinusoidal driving (eq.~\ref{eq:work:sinus}). 
From those equations and the afore mentioned experimental observation on the fluctuations
$\delta \theta$, we see that the fluctuations $\delta W_\tau$
have a gaussian distribution, so the PDF of the work $W_\tau$
done by the external torque is fully characterized by its mean 
${\bar W}_\tau$ and variance $\sigma^2 = \left\langle (\delta W_\tau)^2 \right\rangle$
where $\langle.\rangle$ stands for ensemble average. 
In such a case the FTs take a simple form:
\begin{equation}
S(W_\tau)= {2 \bar{W} \over \sigma^2} W_\tau = \Sigma(\tau) W_\tau, 
\label{eqFTs}
\end{equation}
where $\Sigma(\tau)\equiv(1-\epsilon(\tau))^{-1}$
following the notation of ref.\cite{Cohen}.
From eq.\ref{eqFTs} it is clear that to estimate the finite time correction 
$\Sigma(\tau)$ we need only to compute $\bar{W_\tau}$ and $\sigma^2$. 

$\bar W_\tau$ is simply obtained by inserting in eq.\ref{barW} 
the expression of $M(t)$ and the solution $\bar\theta$ of 
eq.\ref{eqoscillator} with $\eta=0$. The variance $\sigma^2$
is computed using eq.\ref{eqdeltaW}. For the linear ramp of 
fig.\ref{driver_fluct}, we obtain:
\begin{eqnarray}
\sigma^2 &=& \frac{1}{(k_BT)^2}	\frac{M_0^2}{\tau_r^2} 
	\left[ \tau^2 \langle \delta\theta^2 (\tau) \rangle
    + \langle \left( \int_0^\tau \delta\theta(t) dt
    \right)^2 \rangle \right. \nonumber \\
    &-& \left. 2 \tau \int_0^\tau \langle \delta\theta(\tau) \delta\theta(t) \rangle
    dt \right].
    \label{GCL12}
\end{eqnarray}
which can be computed using the correlation
function $R(\tau)=\langle\delta \theta(\tau) \delta\theta(0)\rangle$. 
As already explained, the experimental data indicate that the statistical
properties of $\delta\theta(t)$ on the ramp are the same as the properties of the equilibrium
fluctuations, which are well described by a second order Langevin dynamics. 
Thus we can use for $R(\tau)$ the known equilibrium correlation function of the
thermal fluctuations which for a second order Langevin dynamics
is~\cite{DouarcheJSM}:
\begin{equation}
R(\tau)= \frac{k_BT}{C} {\sin{(\psi|\tau| + \varphi)} \over \sin{\varphi}} \exp{(-\alpha |\tau|)}\,,
\label{correlation2}
\end{equation}
where
$\alpha = 1/\tau_\alpha$, 
$\alpha^2 + \psi^2 = \omega_0^2 = {C / I_{\mathrm{eff}}}$ and  
$\sin{\varphi} = \psi/\omega_0$.
We checked our method on a first order Langevin equation, for
which the exact results of refs.\cite{Cohen,Cohen1} are available.
We find that $\epsilon(\tau)$ for the work computed with our
technique in a first order Langevin dynamics is the same as in
ref.\cite{Cohen,Cohen1} both for TFT and SSFT. Thus
we can now safely apply our technique to a second order Langevin equation.
We find that in the case of the TFT $\epsilon=0\ \forall \tau$, whereas in
the case of the SSFT
\begin{eqnarray}
\epsilon(\tau) = \frac{2}{\psi\tau} \left\{
	\frac{\sin3\varphi}{\omega_0 \tau} 
	- {\exp(-\alpha\tau)} \times \nonumber \right.\\
\left. \left( {\sin(2\varphi+\psi\tau)} 
 	+ \frac{\sin(3\varphi+\psi\tau)}{\omega_0 \tau} 
\right) \right\} \,.
%
%
\label{eqSigmaSSFT}
\end{eqnarray}
The same calculations can be performed for any kind of $M(t)$.
For example with a sinusoidal forcing $M(t)=M_o \sin(\omega_d t)$, 
SSFT for the work $W_n$ defined in eq.~\ref{eq:work:sinus} gives
\begin{eqnarray}
\epsilon (\tau_n) &=& -\frac{\cos2\gamma}{2\alpha\tau_n} \ \frac{\omega_0^2 + \omega_d^2}{\omega_d^2}
+ \frac{1}{\tau_n}{\cal O} \left({\rm e}^{-\alpha\tau_n}\right)
\label{eqepSINUS}
\end{eqnarray}
where $\gamma$ is the phase shift between $\bar{\theta}(t)$ and $M(t)$,
{\em i.e}, $\tan(\gamma) = - 2 \alpha \omega_d / (\omega_0^2 - \omega_d^2)$ 
and $\tau_n=2 n \pi/\omega_d$ with $n$ integer. In Eq.~\ref{eqepSINUS},
${\cal O} \left({\rm e}^{-\alpha\tau_n}\right)$ is a term that 
vanishes exponentially in $\alpha\tau_n$, the expression of which 
is complicated and will be reported in a longer article, together 
with many other interesting features.

These analytical results agree remarkably well with the
experimental results for TFT and SSFT for the work fluctuations in
a harmonic oscillator (see Figs.~\ref{fig:SSFT}c and \ref{fig:sinus}e).

In conclusion we have applied the FTs to the  work fluctuations of
an oscillator driven out of equilibrium by an external force. The
TFT holds for any time whereas the SSFT presents a complex
convergence to the asymptotic behavior which strongly depends on
the form of the driving. The exact formula of this convergence can
be computed using several experimental evidences of the statistics
of the fluctuation. These results are useful for many applications
going from biological systems to nanotechnology, where the
harmonic oscillator is the simplest building block.

This work has been partially supported by EEC contract DYGLAGEMM.


\vfill
\begin{thebibliography}{0}

\bibitem{Evansetal93}
D.~J.\ Evans, E.~G.~D.\ Cohen, and G.~P.\ Morriss, {\em Phys.\ Rev.\ Lett.} \textbf{71}, 2401 (1993);
D.~J.\ Evans and D.~J.\ Searles, {\em Phys.\ Rev.\ E} \textbf{50}, 1645 (1994).

\bibitem{GallavottiCohen95a}
G.~Gallavotti and E.~G.~D.\ Cohen, {\em Phys.\ Rev.\ Lett.} \textbf{74}, 2694 (1995); 
E.~G.~D.\ Cohen, {\em Physica} \textbf{240}, 43 (1997); 
E.~G.~D.\ Cohen and G.~Gallavotti, {\em J.\ Stat.\ Phys.} \textbf{96}, 1343 (1999).

\bibitem{Kurchan98}
J.~Kurchan, {em J.\ Phys.\ A, Math.\ Gen.} \textbf{31}, 3719 (1998);
J.~L.\ Lebowitz and H.~Spohn, {\em J.\ Stat.\ Phys.} \textbf{95}, 333 (1999).

\bibitem{Farago}
J. Farago, {\em J. Stat. Phys.} {\bf 107} 781 (2002);
{\em Physica A} {\bf 331} 69 (2004).

\bibitem{Cohen1} 
R. van Zon and E.G.D. Cohen, {\em Phys. Rev. Lett.} {\bf 91} (11) 110601 (2003);
{\em Phys. Rev. E} {\bf 67} 046102 (2003).

\bibitem{Cohen} 
R. van Zon, S. Ciliberto, E.G.D. Cohen, {\em Phys. Rev. Lett.} {\bf 92} (13) 130601 (2004).

\bibitem{Evans:Searles}
D.J. Evans, D.J. Searles, {\em Advances in Physics} {\bf 51} (7) 1529 (2002).

\bibitem{otherexperiment}
S. Ciliberto and C.~Laroche, {\em J.\ Phys.\ IV}, France {\bf 8}, 215 (1998); 
S. Ciliberto, N. Garnier, S. Hernandez, C. Lacpatia J.-F. Pinton, G. Ruiz Chavarria, {\em Physica A} {\bf 340}, 240 (2004); 
K. Feitosa, N. Menon, {\em Phys.\ Rev.\ Lett.} {\bf 92} 164301 (2004).

\bibitem{Wangetal:02:05}
G.M. Wang, E.M. Sevick, E. Mittag, D.J. Searles, D.J. Evans, {\em Phys.\ Rev.\ Lett.} {\bf 89} 050601 (2002); 
G.M. Wang, J.C. Reid, D.M. Carberry, D.R.M. Williams, E.M. Sevick, D.J. Evans, {\em Phys.\ Rev.\ E} {\bf 71} 046142 (2005).

\bibitem{Garnier}
N. Garnier, S. Ciliberto, {\em Phys.\ Rev.\ E} {\bf 71} 060101(R) (2005).

\bibitem{Zamponi:05}
F. Zamponi, F. Bonetto, L. Cugliandolo, J. Kurchan, {\em J. Stat. Mech.} p09013 (2005).

%

\bibitem{rsi}
F. Douarche, L. Buisson, S. Ciliberto, A. Petrosyan, {\em Rev. Sci. Instr.} {\bf 75} (12) 5084 (2004).

\bibitem{DouarcheJSM}
F. Douarche, S. Ciliberto, A. Petrosyan, {\em J. Stat. Mech.} p09011 (2005).


\end{thebibliography}
\end{document}